# EFFECT OF SECONDARY ECHO SIGNALS IN SPIN-SYSTEMS WITH A LARGE INHOMOGENEOUS BROADENING OF NMR LINE

**J.G.Chigvinadze, G.I.Mamniashvili, Yu.G.Sharimanov**

*©Andronikashvili Institute of Physics*

*Georgia, 38007, Tbilisi, Tamarashvili 6*

## ABSTRACT

The possibility of comparatively simple and fast determination of characteristic relaxation parameters $T_1$, $T_2$ and $T_3$ for nuclear spin-systems with strong Larmor and Rabi inhomogeneous broadenings of NMR lines using the secondary echo signal effect was experimentally shown. Resides, this method gives opportunity to obtain a valuable infomation on the inhomogeneous NMR broadening which reflects the character of magnetic field microscopic distribution in such systems, as example, multidomain magnetics and superconductors.

Spin-echo phenomenon discovered by Hahn in 1950 [1] is now one of the most developed method of the nuclear magnetic resonance (NMR) and the electronic paramagnetic resonance effectively used for studying of physical properties of very different materials as metals, dielectrics  semiconductors and so on.

It is known that for observation of the Hahn echo at least two radio frequency (RF) pulses are necessary. It is also known that information on spin-system usually obtained with the help of two pulse Hahn echo (TPE) could be obtained by the single-pulse echo (SPE) [2,3]. But in a number of cases the most convenient proves to be the spin echo excitation by multipulse series of RF pulses which, in particular, provide an operative measurement of relaxation times $T_1$ and $T_2$  of the investigated systems [4].

Recently in [5,4] the formation of secondary nuclear spin echo signals in systems with large Larmor and Rabi inhomogeneous broadenings of a NMR line at periodical excitation of a sample by series of single or two RF pulses was considered in conditions when the spin-system unequilibrium preceded RF pulses .

The important examples of such systems represent nuclei arrainged in domain walls (DW) of magnetics and in normal cores of Abrikosov`s vortex in superconductors of the second type (Sc.II).

Earlier similar phenomena in more simple systems, as protons in water solutions of paramagnetic ions and aluminates, were studied in works [6,8].

As it is known in such systems phenomenological and quantum theories of spin echo [6,9] predict generation only a single echo signal at application on spin-system series of two RF pulses.

In work [7]  it was pointed out a possibility of  generation of additional echo signals, if unequilibrium of spin-system preceded the excitation by RF pulse, and in [8] the theoretical explanation was given and a comparison with the experiment  was made.

By the way a conclusion was made that properties of main and secondary echoes, obtained at a periodical excitation by series of pulses, give the new possibilities of experimental determination of the fundamental characteristic times $T_1$, $T_2$ and $T_3$ of spin-systems.

From this point of view the investigation of the effect of secondary echo signals  at excitation by periodical series in the case of systems with large Larmor and Rabi inhomogeneous broadenings could also provide a new possibility of investigation of the nuclear spin-system dynamics in magnetics and type II superconductors, including high-temperature superconductors (HTSC).  The investigation of this possibility is aimed in this work.

In work [5], using the statistical tensor formalism, a theoretical investigation was made on formation of SPE and its secondary signals at the presence of both Larmor and Rabi inhomogeneous broadenings of NMR lines, at conditions when the period T of RF pulse repetition obeys inequality $T_3 \ll T_2 < T < T_1$, and only the longitudinal component of the nuclear magnetization is important before RF pulse, and an unequilibrium of spin-system precedes RF pulses .It was shown in [5] that a dephasing of nuclear spin-system was accumulated during n pulse excitations and restored within a time interval elapsing from the trailing edge of the last "counting" [(n+1)th] pulse of a multipulse train. It is on this basis that the evolution away from OFID (LIB only) or echolike step anomaly (LIB plus RIB) for a single-pulse excitation, toward the formation of a well-defined echo at $\tau$ ($\tau$ is a RF pulse duration), together with the emergence of secondary echoes for a multipulse train was established [5] analytically and numerically.

Let us note that in conditions of work [5] similar expressions for nuclear magnetization vectors could be obtained following the usual classical approach by solution of Bloch equations [10]. Besides this, the above pointed methods could be applied also in the case of a periodical two-pulse excitation. It was also shown in [5] that secondary echoes formation effect is present for the large Larmor inhomogeneity in isolation, but is stronger at the simultaneous presence of both frequency inhomogeneities, a characteristic of multidomain ferromagnets.

Let us illustrate some of the above pointed problems on examples of a practical interest.

Experimental results were obtained by the NMR spectrometer Bruker "Minispec p20" provided with the signal averager "Kawasaki Electronica" at room temperatures.

Niobium hydride could be considered as an example of systems with both types of inhomogeneities of the NMR line. In this case the Rabi frequency inhomogeneity is a result of the skin-effect.

Fig.1 shows a signal averager record presenting proton echo signals in niobium hydride at conditions of an unequilibrium of spin-system at the periodical excitation.

Let us consider in more details a secondary two-pulse echo signals on the example of protons in silicon oil (Silicone KF 96-type), Fig.2, and in silicon oil (SO) mixed with a powder sample of $YBa_2Cu_3O_7$ (SO+YBCO), which is the object similar the one used in [11] for investigation of inhomogeneous broadening of the NMR line in HTSC due to the formation of Abrikosov vortex lattice, (Fig.3).

Fig.4 shows the primary and secondary echo signal dependences on repetition period T of pairs of RF pulses It could be seen from Fig.4 that the maximum of secondary echo on T takes place at T~0.5 $T_1$, where $T_1$ is a value of spin-lattice relaxation time obtained by the known inverting 180°-90° sequence method [4]. and at the same time a decay factor of the secondary echo intensity dependence on T is close to the value of $T_1$ obtained by the 180°-90° method. So the measurement of the secondary echo intensity dependence on T gives a relatively simple method of determination in one passage the spin-lattice relaxation time $T_1$ without a further elaboration of experimental data .

It is seen also that at a phase sensitive detection the phase of secondary echo signal is opposite to the one of the first echo signal (Fig.5). This effect was theoretically obtained in [5] for the secondary signal of single pulse echo but was not observed experimentally.

Let us note that observed secondary echo signals are transient coherences that arise from a complex superposition of oscillatory free induction decays, which extend further into the time domain because of cumulative dephasing within each pulse in a multipulse sequence. It gives us a possibility to investigate the character of inhomogeneous broadening of NMR line without constraints connected with a "dead time" of receiver .Besides this a comparison of experimental data with results of numerical modeling of the NMR line shape using the nonlinear nuclear spin-dynamics equations [5] allows one principally to obtain valuable information on objects being studied, as example, the distribution function of RF gain factors in multidomain magnetics, and also on the microscopical character of magnetic field changes in HTSC ,in normal metals, and so on.

In conclusion, It is shown that in the frames of the described approach, besides the comparatively simple possibility of obtaining information on relaxation parameters $T_1$, $T_2$, and $T_3$ using the same two-pulse RF train, there is also possibility of obtaining the detailed information on the nuclear spin-system dynamics in multidomain magnetics and HTSC that deserves further investigations.

This work is made in the frames of ISTC Project G-389.

FIGURE CAPTIONS

Fig.1 Record of primary and secondary proton echo signals in niobium hydride $NbH_{0.68}$ .

     Marks show the time position of RF pulses.

Fig.2 Multiple proton echo signals in silicon oil (SO) at T=300K.

Fig.3 Multiple proton echo signals in SO mixed with YBCO powder, (SO + YBCO).

Fig.4 Intensity I dependences of primary (1) and secondary (2) echo signals on repetition

     period T of two exciting RF pulses T  in SO + YBCO.

Fig.5 Primary and secondary echoes at a phase-sensitive detection in SO + YBCO.

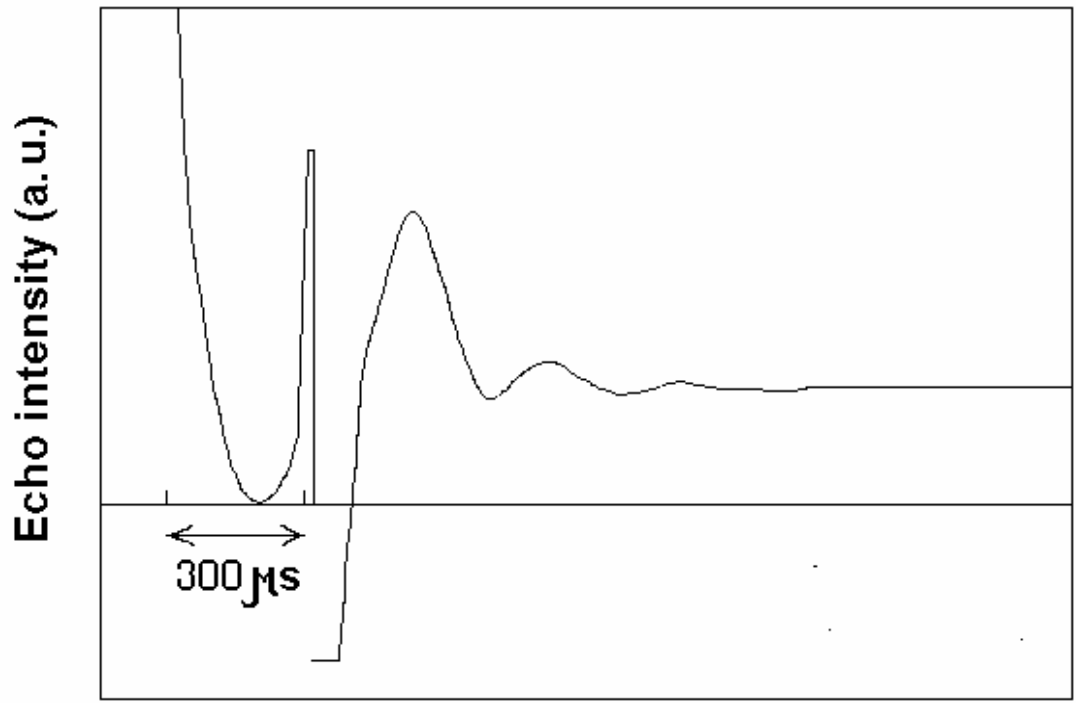

**Fig.1**

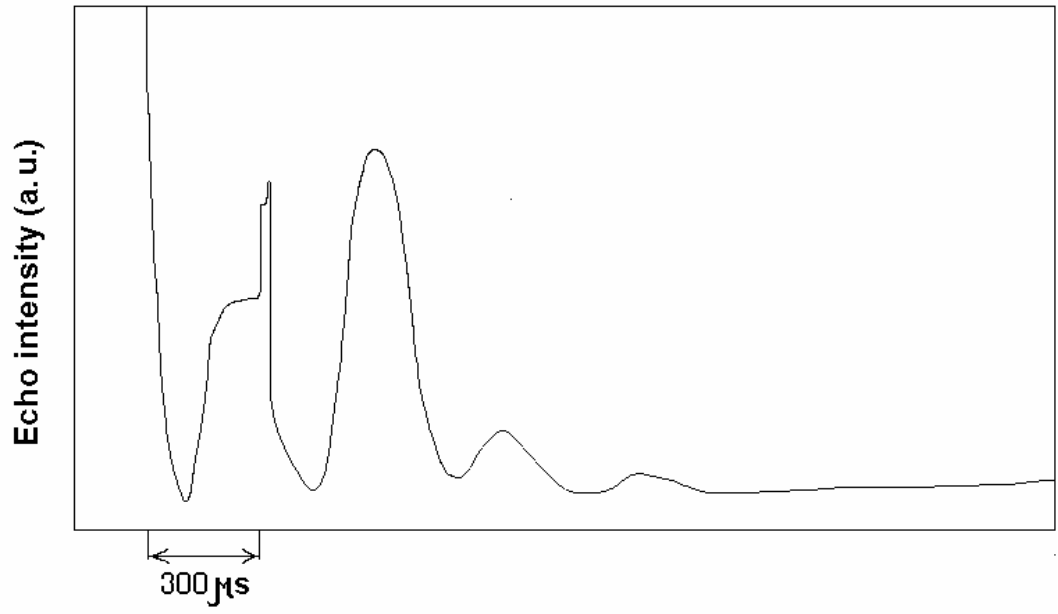

**Fig.2**

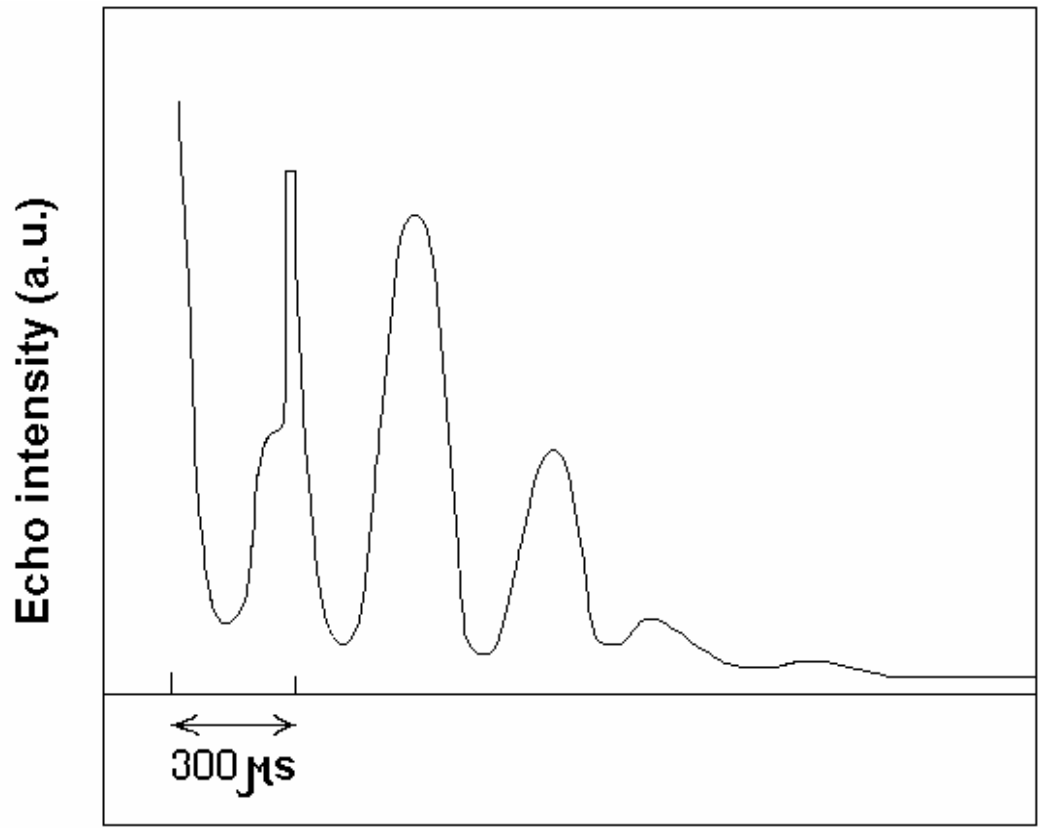

**Fig.3**

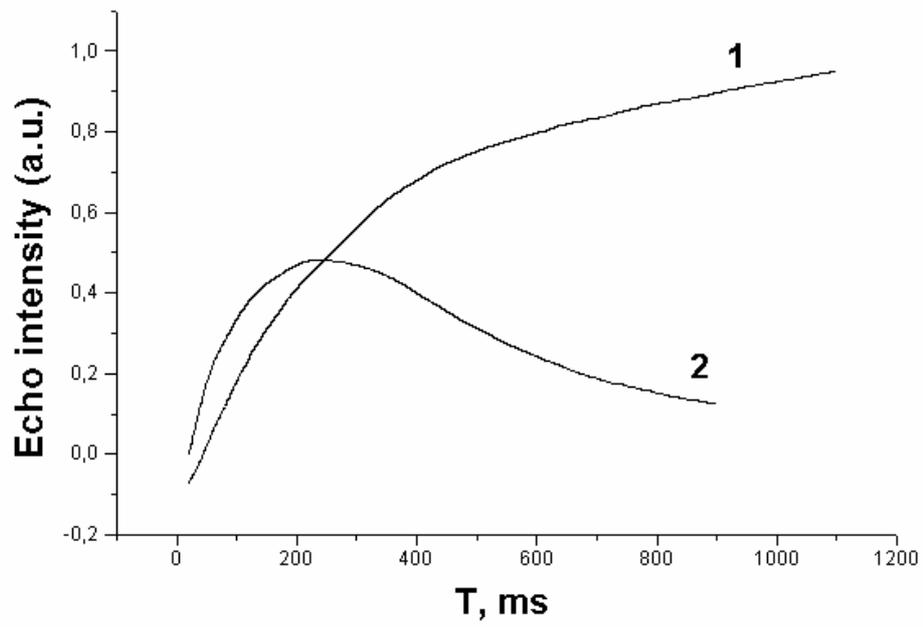

**Fig.4**

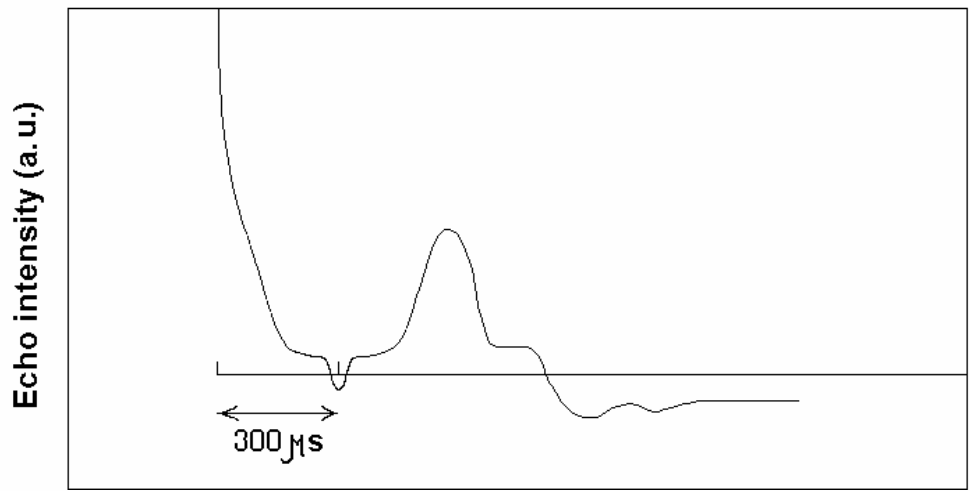

**Fig.5**